\documentclass[lettersize,journal]{IEEEtran}
\usepackage{algorithm}
\usepackage{algpseudocode}
\usepackage{xcolor}
\usepackage{comment}
\usepackage[nocompress]{cite}
\def\BibTeX{{\rm B\kern-.05em{\sc i\kern-.025em b}\kern-.08em
    T\kern-.1667em\lower.7ex\hbox{E}\kern-.125emX}}

\usepackage{amsbsy,amsmath,amssymb,amsfonts,amsthm}
\usepackage{bm}
\usepackage{gensymb}
\usepackage{graphicx,wrapfig}
\usepackage{multirow}
\usepackage{booktabs}
\usepackage{mathrsfs}
\usepackage{color, soul}

\usepackage[hidelinks]{hyperref}
\usepackage{cite}
\usepackage[shortlabels]{enumitem}
\usepackage{epsfig}
\usepackage{epigraph}
\usepackage{csquotes}
\usepackage[strict]{changepage}
\usepackage{datetime2}
\usepackage{siunitx}
\usepackage{tabularx}
\usepackage{acronym}
\usepackage{orcidlink}

\usepackage{import}
\usepackage{xifthen}
\usepackage{pdfpages}
\usepackage{transparent}
\usepackage{pstricks}
\usepackage[inkscapearea=page]{svg}
\svgpath{{./graphics/}}

\usepackage{float}
\usepackage{tikz}
\usepackage{pgfplots}
\DeclareUnicodeCharacter{2212}{−}
\usepgfplotslibrary{groupplots,dateplot}
\usetikzlibrary{patterns,shapes.arrows,shapes.geometric,arrows,pgfplots.groupplots,positioning,arrows.meta}
\pgfplotsset{compat=newest}

\renewcommand{\rm}[1]{\mathrm{#1}} 
\newcommand{\bs}[1]{\boldsymbol{#1}} 
\renewcommand{\bf}[1]{\mathbf{#1}} 
\newcommand{\uli}[1]{\underline{#1}} 
\newcommand{\oli}[1]{\overline{#1}} 

\newcommand{\pb}[4][black]{%
  \parbox{#3cm}{%
    \ifthenelse{\equal{#2}{l}}{\raggedright}{%
    \ifthenelse{\equal{#2}{c}}{\centering}{%
    \ifthenelse{\equal{#2}{r}}{\raggedleft}{}}}%
    \textcolor{#1}{#4}%
  }%
}

\newcolumntype{H}{>{\setbox0=\hbox\bgroup}c<{\egroup}@{}}

\usepackage{hyperref}

\newif\ifallknn
\allknntrue

\begin{document}

\title{Learning to Fix: Optimisation-Aware Machine Learning for Accelerated Unit Commitment}




\author{Benjamin Fritz\,\orcidlink{0009-0006-5326-8299},~\IEEEmembership{Student Member,~IEEE}, Andreas Makrides\,\orcidlink{0009-0001-1795-0729},~\IEEEmembership{Student Member,~IEEE}, Maryam Fetanat\,\orcidlink{0009-0008-0029-093X},\, Pierre Pinson\,\orcidlink{0000-0002-1480-0282},~\IEEEmembership{Fellow,~IEEE}\vspace{-6mm}}



\maketitle

\begin{abstract}
Unit Commitment is a computationally demanding mixed-integer linear optimisation problem requiring many binary commitment decisions across a scheduling horizon. To reduce this computational burden, machine learning approaches can predict a subset of these decisions, thereby shrinking the search space explored by the optimisation solver. Existing confidence-based approaches determine which variables to fix based on user-defined probability thresholds, which are agnostic to the downstream optimisation problem. We address this limitation by introducing an optimisation-aware framework that yields generator-specific confidence thresholds based on the impact of fixing errors on Unit Commitment solution quality. This approach won first place in the 2025 EPRI AI-ccelerating Unit Commitment competition. The proposed framework achieves a mean optimality gap below 0.5\% while also delivering an average speed-up of more than 20×. It therefore provides a general mechanism for translating probabilistic predictions into controlled reductions of mixed-integer search spaces, directly linking learning decisions to their downstream computational and economic consequences.
\end{abstract}


\section{Introduction}\label{sec:intro}


\IEEEPARstart{U}{nit Commitment (UC)} is a fundamental optimisation problem in power system operations. It determines the on/off status of generating units over a scheduling horizon, together with their dispatch levels, while satisfying demand, reserve requirements, ramping limits, generators' minimum up- and down-times, and other technical operating limits \cite{woodPowerGenerationOperation2013}. The problem is commonly formulated as a mixed-integer linear optimisation problem (MILP), where binary variables represent commitment decisions and continuous variables represent generation levels and other operational quantities \cite{knuevenMixedIntegerProgrammingFormulations2020a, padhyUnitCommitmentaBibliographical2004}. Although commercial solvers exist to solve these problems and have considerably improved the tractability of UC, the problem remains computationally demanding, particularly for large systems, long scheduling horizons, and applications where many instances must be solved within limited time. 

Higher shares of renewable generation, increased participation of storage assets, and tighter operational margins can lead to more volatile and uncertain system conditions, increasing the need to address the computational burden of UC \cite{vanackooijLargescaleUnitCommitment2018}. In market clearing, security assessment, planning studies, and real-time decision-support tools, operators may need to solve these types of problems repeatedly. In these settings, even moderate reductions in solution time can be valuable, provided that feasibility and solution quality are not compromised.

Machine learning (ML) has recently attracted interest as a way to accelerate UC and other similar problems. Existing approaches can be broadly grouped together into two categories \cite{millaUseArtificialIntelligence2025}. The first uses surrogate models to directly predict commitment schedules or complete UC solutions, including dispatch variables. In the context of the UC, the majority of such surrogate models, also referred to as \textit{solver-free} methods, employ Reinforcement Learning, modelling the UC as a Markov Decision Process (e.g., \cite{omalleyReinforcementLearningMixedinteger2023,carrionComputationallyEfficientMixedinteger2006}). These methods can be very fast, but they are likely to produce infeasible or highly suboptimal solutions. To overcome these issues, a second category leverages ML to guide or modify a downstream optimisation algorithm, i.e., \textit{solver-augmented} methods. For the UC, such methods are often used to \textit{(i)} screen out redundant or likely inactive constraints \cite{pinedaDataDrivenScreeningNetwork2020}, \textit{(ii)} predict warm-start hints for the solver \cite{pourahmadiUnitCommitmentPredictor2025}, or \textit{(iii)} tighten the feasible search space of the downstream algorithm \cite{xavierLearningSolveLargeScale2021,sugishitaUseMachineLearning2024}.

In this last realm of learning-assisted model reduction, multiple
works have leveraged \textit{confidence-based variable fixing} to reduce the dimensionality of the binary commitment variable space \cite{linProbabilisticForecastingGenerators2020,xavierLearningSolveLargeScale2021,schmittFastSolutionUnit2022,shekeewMachineLearningAdditionalDecision2023,gaoOnlineLearningStable2023,rameshSpatioTemporalDeepLearningAssisted2024,rameshFeasibilityLayerAided2024,zaterMultiStageWarmStartDeep2026}. Among these, \cite{xavierLearningSolveLargeScale2021,shekeewMachineLearningAdditionalDecision2023} evaluate the offline performance of their trained classifiers to decide if a variable hint can be trusted. For example, if a classifier correctly predicts the commitment state of a generator across all training instances, they will confidently trust the classifier's predictions to fix this variable at inference. Conversely,
\cite{linProbabilisticForecastingGenerators2020,schmittFastSolutionUnit2022,rameshSpatioTemporalDeepLearningAssisted2024,rameshFeasibilityLayerAided2024,zaterMultiStageWarmStartDeep2026} delay the variable fixing decisions until the online stage. They use probabilistic classifiers to decide if a variable should be fixed, given a user-specified confidence threshold. Here, if the classifier predicts a commitment probability higher than, say, 90\%, the associated commitment variable is fixed to ON, otherwise, the decision is deferred to the downstream solver.

In line with the above works, this paper proposes an ML-assisted UC framework based on confidence-based binary fixing using probabilistic classifiers. We first revisit the idea of predefined confidence thresholds that offer no direct control over the cost impact of variable fixing. Conservative choices can limit computational savings, whereas aggressive fixing can increase operating cost or render the reduced UC problem infeasible. We then propose methods to compute individual thresholds for each generator that are tuned by accounting for the effect of fixing errors on the downstream UC objective value. This allows the framework to distinguish between misclassifications that have little impact on operating cost and those that may lead to larger suboptimality or infeasibility. A key difficulty in this setting is that the UC problem is a MILP. Embedding continuous convex optimisation problems into ML models has been successful in part because such problems can be represented through differentiable optimisation layers \cite{amosOptNetDifferentiableOptimization2017}. In contrast, MILPs contain discrete variables and are therefore not directly differentiable, which makes it difficult to propagate cost- or feasibility-based information through the learning model. The proposed framework addresses this issue by keeping the supervised learning task separate from the optimisation-aware calibration step. The classifier is trained to produce probabilistic commitment predictions, while the cost- and constraint-related information of the UC problem is introduced later through the threshold-tuning procedure. This differs from approaches that attempt to incorporate optimisation structure directly into the learning task \cite{ferberMIPaaLMixedInteger2020,tangLearningOptimizeMixedInteger2025}.



The main contributions of this paper are summarised as follows:
\begin{itemize}
    \item We propose an optimisation-aware threshold-selection method that accounts for the effects of binary variable fixing on UC feasibility and operating cost. A decomposition algorithm determines generator-specific confidence thresholds subject to a prescribed cost tolerance on validation instances. This tolerance controls the trade-off between problem reduction and solution quality while keeping threshold tuning separate from classifier training.
    \item We evaluate the method on the \textit{AI-ccelerating Unit Commitment} competition dataset against the full MILP benchmark and existing learning-based approaches. Using a $k$-nearest-neighbour ($k$NN) classifier and a 1\% validation cost tolerance, the method achieves an average speed-up of 20.8, an average optimality gap of 0.48\%, and feasible solutions for 99.81\% of test instances. Complementary results using the CatBoost architecture employed in the competition are reported in Appendix~\ref{app:catboost}. Together, these results demonstrate that tuning the fixing thresholds can deliver substantial computational savings across different classifier architectures.
\end{itemize}

The remainder of this paper is organised as follows. Section~\ref{sec:uc_problem} presents the UC formulation. Section~\ref{sec:proposed_methodology} describes the variable-fixing strategies and the proposed threshold-tuning framework. Section~\ref{sec:exp_study_results} presents the experimental setup and results. Section~\ref{sec:conclusions} concludes the paper and discusses potential extensions.


\section{Unit Commitment Problem Formulation}
\label{sec:uc_problem}

The UC problem determines the scheduling and dispatch of generating units over a finite time horizon \cite{woodPowerGenerationOperation2013}. Let $\mathcal{G}$ and $\mathcal{T}$ denote the sets of generating
units and time periods, indexed by $g$ and $t$, respectively. The main decision variables are the binary commitment status $u_{gt} \in \{0,1\}$ and the continuous power output $p_{gt} \in \mathbb{R}_{\geq 0}$. The output of each unit is subject to unit-dependent generation limits,
together with additional technical constraints (e.g., ramp rates).  For ease of notation, we further define the following vectors: $\bf{p}_{g}=\{p_{gt}\}_{t\in\mathcal{T}}$ and $\bf{u}_{g}=\{u_{gt}\}_{t\in\mathcal{T}}$, as well as $\bf{P}=\{\bf{p}_{g}\}_{g\in\mathcal{G}}$ and $\bf{U}=\{\bf{u}_{g}\}_{g\in\mathcal{G}}$.

The problem is formulated as a MILP that minimises total operating cost, while meeting the load $L_t$ at all times:
\begin{subequations}\label{eq:UC}
    \begin{alignat}{2}
    C^*\,=\,\min_{\bf{P},\bf{U}} \quad & \sum_{g \in \mathcal{G}} c_g(\bf{p}_g,\bf{u}_g) \\
    \text{s.t.} \quad &
    \sum_{g \in \mathcal{G}} p_{gt} = L_t \quad &\forall t \in\mathcal{T} ,\label{seq:p_balance}\\
    &
    \left(\bf{p}_g,\bf{u}_g\right) \in \Pi_g
    \quad &\forall g \in \mathcal{G}. \label{seq:g_polytope}
    \end{alignat}
\end{subequations}

Here, $c_g(\mathbf{p}_g,\mathbf{u}_g)$ denotes the generator-specific, linear operating cost function, including production, no-load, start-up, and shut-down cost components. The system is represented using a copper-plate model, where network constraints are not explicitly considered and the main system-level constraint is the balance between aggregate demand $L_t$ and total generation, enforced through \eqref{seq:p_balance}. Without loss of generality, we assume that load shedding decisions can be modelled as a dispatch variable $p_{gt}$, with a high-penalty cost.

The set $\Pi_g$ in \eqref{seq:g_polytope} denotes the generator dispatch polytope, which represents the feasible operating region of generator $g$ for a given commitment trajectory. It includes unit-specific constraints linking commitment and dispatch decisions, such as generation limits and ramping limits and further models the feasible commitment logic, including start-up and shut-down logic and minimum up- and down-time constraints. 
In this study, the default formulation of the \textit{UnitCommitment.jl} library is used. The detailed expressions for the cost function $c_g$ and the constraints defining the generator polytope $\Pi_g$ are therefore omitted here and can be found in the official documentation of the library \cite{santosxavierUnitCommitmentjlJuliaJuMP2025}. 

For later use, let $\bs{\Theta}^{(n)}=\left\{C^{*(n)},\bigl\{L^{(n)}_t\bigr\}_{t\in\mathcal{T}},\bigl\{\Pi_g^{(n)}\bigr\}_{g\in\mathcal{G}}\right\}$, denote a set of parameters describing the UC instance $n$, namely its optimal cost, load profile and generator polytopes. 






\section{Fixing Binary Variables using Probabilistic Classification}
\label{sec:proposed_methodology}

This section presents variable-fixing strategies for the UC formulation in Section~\ref{sec:uc_problem}. Standard binary classification fixes all commitment variables using a single threshold, whereas confidence-based fixing leaves predictions within the two thresholds to the downstream solver. We consider three threshold-selection approaches: predefined, worst-case misprediction, and suboptimality-constrained thresholds. The proposed suboptimality-constrained approach incorporates UC costs and constraints into threshold tuning after classifier training.

\subsection{Standard Binary Classification}\label{ssec:class_standard}

The main complexity in solving problem \eqref{eq:UC} comes from the binary commitment variables $\bf{U}$. If all these variables were known and fixed to an optimal commitment schedule $\bf{U}^*$, the remaining dispatch problem could be solved more efficiently as a linear program (LP), yielding the corresponding optimal dispatch $\bf{P}^*$.  

Mathematically, we can describe such an approach as follows. Assume we have access to a dataset $\bigl\{\bf{z}^{(i)},\bf{U}^{*(i)},C^{*(i)}\bigr\}_{i\in[1:N^{\text{train}}]}$, where $\bf{z}^{(i)}\in\mathcal{Z}$ contains the input features associated with UC instance $i$, $\bf{U}^{*(i)}$ is the corresponding optimal commitment schedule, and $C^{*(i)}$ is the optimal objective value. 
The goal is to learn the parameters $\boldsymbol{\theta}$ of a classifier $h_{\boldsymbol{\theta}}:\mathcal{Z}\rightarrow\{0,1\}^{|\mathcal{G}|\times|\mathcal{T}|}$, which maps an unseen input $\mathbf{z}$ to a predicted commitment schedule, $\hat{\bf{U}}$:
\begin{equation}
    \hat{\bf{U}}= h_{\boldsymbol{\theta}}({\bf{z}}).
\end{equation}

In practice, most ML classifiers do not directly output binary values in $\{0,1\}$. Instead, they are parameterised through a continuous scoring function $f_{\boldsymbol{\theta}}:\mathcal{Z}\rightarrow\mathbb{R}^{|\mathcal{G}|\times|\mathcal{T}|}$ whose outputs are transformed into probabilities using an activation function. For a sigmoid activation function $\sigma(x)=\frac{1}{(1+e^{-x})}$, the predicted probability is given by:
\begin{equation}
    \pi_{gt}({\bf{z}})=\sigma\left([f_{\boldsymbol{\theta}}({\bf{z}})]_{gt}\right) \approx \mathbb{P}\left({u}_{gt}^{*} = 1 \mid {\bf{z}}\right),
\end{equation}
where $u_{gt}^{*}$ denotes the reference commitment decision and $\pi_{gt}({\bf{z}})$ is interpreted as the probability that the unit $g$ is committed at time $t$. For the $k$NN classifier used in this study, commitment probabilities are obtained through inverse-distance weighting of neighbouring training labels, as described in Section~\ref{ssec:ml_architecture}.
In standard binary classification, the predicted commitment status is obtained by applying a fixed threshold $\tau$ to the predicted probability:
\begin{equation}
\hat{u}_{gt} =
\begin{cases}
1, & \text{if } \pi_{gt}({\bf{z}}) \geq \tau, \\ 0, & \text{otherwise}
\end{cases}
\qquad \forall g\in\mathcal{G}, t\in\mathcal{T}.
\end{equation}
A common choice is $\tau=0.5$, although other values may be used depending on the desired trade-off between false positive and false negative commitment decisions.


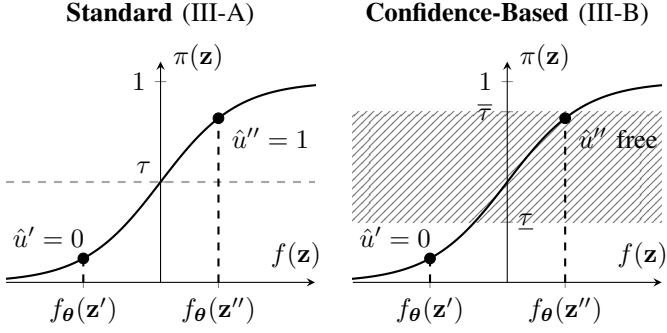
\begin{figure}
    \centering
    \begin{tikzpicture}
\begin{groupplot}[
group style={
group size=2 by 1,
horizontal sep=0.5cm
},
width=0.64\linewidth,
height=4.5cm,
xmin=-4,xmax=4,
ymin=0,ymax=1.1,
set layers,
axis lines=middle,
title style={yshift=4pt},
xtick={-2,1.5},
xticklabels={$f_{\boldsymbol{\theta}}(\bf{z}')$,$f_{\boldsymbol{\theta}}(\bf{z}'')$},
xlabel={$f(\mathbf{z})$}, xlabel style={at={(axis description cs:1.05,0.02)}},
ylabel={$\pi(\mathbf{z})$}, ylabel style={at={(axis description cs:0.51,1.12)}}
]
\nextgroupplot[
title={\textbf{Standard} (\ref{ssec:class_standard})},
ytick={0,0.5,1},
yticklabels={$0$,,$1$}
]
\addplot[thick,domain=-6:6,samples=200]
{1/(1+exp(-x))};
\node[anchor=south east] at (axis cs:0.0,0.5) {$\tau$};
\addplot[dashed,thick,gray!70] coordinates {(-4,0.5) (4,0.5)};
\addplot[mark=*,mark size=2pt,only marks] coordinates {(-2,{1/(1+exp(2))})};
\addplot[dashed,thick] coordinates {(-2,0) (-2,{1/(1+exp(2))})};
\node[anchor=south] at (axis cs:-2.9,{1/(1+exp(2))+0.01}) {$\hat{u}'=0$};
\addplot[mark=*,mark size=2pt,only marks] coordinates {(1.5,{1/(1+exp(-1.5))})};
\addplot[dashed,thick] coordinates {(1.5,0) (1.5,{1/(1+exp(-1.5))})};
\node[anchor=north west] at (axis cs:1.6,{1/(1+exp(-1.5))}) {$\hat{u}''=1$};

\nextgroupplot[
title={\textbf{Confidence-Based} (\ref{ssec:class_confidence})},
ytick={0,0.3,0.85,1},
yticklabels={$0$,,$\oli{\tau}$,$1$},
]
\addplot[draw=none] coordinates {(0,0)};
\node[anchor=west] at (axis cs:0.05,0.3) {$\uli{\tau}$};
\addplot[
draw=none,
fill={rgb,255:red,211;green,211;blue,211},
pattern=north east lines,
pattern color={rgb,255:red,120;green,120;blue,120},
forget plot,
on layer=axis background
]
coordinates {(-4,0.3) (4,0.3) (4,0.85) (-4,0.85)};
\addplot[thick,domain=-6:6,samples=200]
{1/(1+exp(-x))};
\addplot[mark=*,mark size=2pt,only marks] coordinates {(-2,{1/(1+exp(2))})};
\addplot[dashed,thick] coordinates {(-2,0) (-2,{1/(1+exp(2))})};
\node[anchor=south] at (axis cs:-2.9,{1/(1+exp(2))+0.01}) {$\hat{u}'=0$};
\addplot[mark=*,mark size=2pt,only marks] coordinates {(1.5,{1/(1+exp(-1.5))})};
\addplot[dashed,thick] coordinates {(1.5,0) (1.5,{1/(1+exp(-1.5))})};
\node[anchor=north west] at (axis cs:1.6,{1/(1+exp(-1.5))}) {$\hat{u}''$ free};
\end{groupplot}
\end{tikzpicture}
    \vspace{-5mm}
\caption{Left: Single-threshold classification. Right: Confidence-based variable fixing. Predictions outside the confidence thresholds $[\uli{\tau},\oli{\tau}]$ are fixed to binary values, while those inside remain free for the MILP solver to determine.}
    \label{fig:sigmoids}
\end{figure}

\subsection{Confidence-Based Binary Classification}\label{ssec:class_confidence}

The standard classification procedure described above produces a complete binary commitment matrix $\hat{\bf{U}}$, reducing the downstream MILP to an LP. However, this also makes the approach sensitive to classification errors. A single incorrectly fixed commitment variable $\hat{u}_{gt}$ may prohibit the recovery of a near-optimal solution or worse, make the reduced problem infeasible. 

To overcome this issue, a common approach is to use the probability scores $\pi_{gt}$ produced by the classifier to decide which predictions should be fixed and which should remain free. Instead of relying on a single threshold parameter $\tau$, confidence-based fixing uses an interval $[\uli{\tau},\oli{\tau}]$, as illustrated in Figure~\ref{fig:sigmoids}. Predictions outside this interval are treated as sufficiently confident and are fixed, while predictions inside the interval are left to be optimised by the MILP solver. Specifically, for each commitment variable, the fixing rule is defined as:
\begin{equation}
    \hat{u}_{gt} =
    \begin{cases}
    1, & \pi_{gt} > \oli{\tau},\\
    0, & \pi_{gt} < \uli{\tau},\\
    \text{free}, & \pi_{gt} \in [\uli{\tau},\oli{\tau}],
    \end{cases}
    \qquad \forall g\in\mathcal{G}, t\in\mathcal{T}.
    \label{eq:fixing}
\end{equation}

This idea is related to the idea of \textit{classification with a reject option} \cite{herbeiClassificationRejectOption2006,geifmanSelectiveNetDeepNeural2019} in statistics, where uncertain predictions are not forced into one of the available classes. In the optimisation setting, however, rejected predictions are not discarded; they are passed to the optimisation model and decided as part of the reduced UC problem. This is important because uncertain commitment decisions may still be resolved consistently once the full set of UC constraints and cost terms is considered.

A crucial aspect is how to choose the thresholds that define the grey zones. In the following, we present three different approaches for this. The first one assumes that the grey zones are the same for all variables, while the latter two derive generator-specific thresholds $\uli{\tau}_g$ and $\oli{\tau}_g$. In theory, one could even aim to find specific thresholds for unit and timestep, i.e., $\uli{\tau}_{gt}$ and $\oli{\tau}_{gt}$. However, for simplicity and improved robustness through temporal consistency, this approach was not considered here. We also assume that a dataset of $N^{\rm{val}}$ unseen instances has been set aside before training. This validation dataset will later be used for tuning the thresholds in the latter two methods.

\subsubsection{Constant Thresholds}\label{sssec:const_thresholds}

This is the predominant approach used in the literature. Here, the researchers define a constant percentage $r_{\%}$, e.g., 1\% in \cite{schmittFastSolutionUnit2022}, 10\% in \cite{rameshFeasibilityLayerAided2024}, that is shaved off from the bottom and top of the interval $[0,1]$, i.e.,
\begin{equation}
    \uli{\tau}_g=\frac{r_{\%}}{100\%}\quad,\quad \oli{\tau}_g=1-\frac{r_{\%}}{100\%}\quad\forall g \in\mathcal{G}.
\end{equation}
However, such an approach does not account for the fact that different classifiers may have different accuracy. The following two methods propose ideas on how to incorporate this heterogeneity among the classification tasks.

\subsubsection{Worst-Case Misprediction Thresholds}\label{sssec:worst_case}
A simple and intuitive way to construct the confidence thresholds is to use the worst-case probability scores observed on the validation set for optimal on- and off-commitment decisions. For each generator $g$, the lower and upper thresholds are defined as: 
\begin{subequations}
\begin{equation}
    \uli{\tau}_g=\min_{t\in\mathcal{T}}\min_{i\in{[N^{\text{val}}]}}\left\{\pi_{gt}\left(\bf{z}^{(i)}\right)|\,u^{*(i)}_{gt} = 1\right\}\quad\forall g \in\mathcal{G},
\end{equation}
\begin{equation}
    \oli{\tau}_g=\max_{t\in\mathcal{T}}\max_{i\in{[N^{\text{val}}]}}\left\{\pi_{gt}\left(\bf{z}^{(i)}\right)|\,u^{*(i)}_{gt} = 0\right\}\quad\forall g \in\mathcal{G}.
\end{equation}
\end{subequations}
With this approach, the $[\uli{\tau}_g,\oli{\tau}_g]$ intervals contain the range of probability scores where validation samples with different true commitment statuses overlap. If $\oli{\tau}_g<\uli{\tau}_g$, the two threshold values do not define an overlapping interval. In this case, we set both of them equal to their midpoint $\frac{(\oli{\tau}_g+\uli{\tau}_g)}{2}$. 

The main limitation of this approach is that it can result in rather conservative thresholds, as they are constructed from specific optimal schedules and may be driven by difficult or atypical samples. Further, all misclassifications are treated equally, although not every deviation has the same impact on the optimality and feasibility of the solution.

\subsubsection{Suboptimality-Constrained Thresholds}\label{sssec:suboptimal}

A more flexible approach is to make the confidence thresholds depend on the impact of the corresponding fixing decision. If a misclassification has only a marginal effect on the objective value, the interval $[\uli{\tau}_g,\oli{\tau}_g]$, can be kept tight, allowing the variable to be fixed more often. In contrast, for riskier decisions that may significantly affect cost or feasibility, the thresholds should be wider, leaving the variable free to be optimised by the solver.

Assume that the user specifies an upper-bound $\varepsilon^{\max}_{\%}$ on the allowable suboptimality gap of the downstream UC problem. Further, let $\uli{\bs{\tau}} = \{\uli{\tau}_g\}_{g\in\mathcal{G}}$ and $\oli{\bs{\tau}} = \{\oli{\tau}_g\}_{g\in\mathcal{G}}$. For a single instance $i$, the tightest thresholds that satisfy this suboptimality limit are obtained by solving the following MILP:
\begin{subequations}\label{eq:tau_tuning}
    \begin{align}
    \mkern-15mu\min_{\substack{\uli{\bs{\tau}},\oli{\bs{\tau}},\\\bf{P}^{(i)},\bf{U}^{(i)}}} \quad & d\left(\uli{\bs{\tau}},\oli{\bs{\tau}}\right)\\[1mm]
    \text{s.t.} \quad & \underline{\smash{\text{Threshold logic}}}
    \notag\\
    & 1-u_{gt}^{(i)}\ge\uli{\tau}_g-\pi_{gt}^{(i)}
    \qquad\forall g\in\mathcal{G},t\in\mathcal{T},
    \label{seq:tau_minus_toggle}\\
    & u_{gt}^{(i)}\ge\pi_{gt}^{(i)}-\oli{\tau}_g
    \qquad\forall g\in\mathcal{G},t\in\mathcal{T},
    \label{seq:tau_plus_toggle}\\
    & 0\le\uli{\tau}_g\le\oli{\tau}_g\le1
    \qquad\forall g\in\mathcal{G},
    \label{seq:pos}\\[2mm]
    & \underline{\smash{\text{Suboptimality limit}}}
    \notag\\
    & \frac{\sum_{g \in \mathcal{G}} c_g\bigl(\bf{p}^{(i)}_g,\bf{u}^{(i)}_g\bigr)-C^{*(i)}}{C^{*(i)}}\le \frac{\varepsilon^{\max}_{\%}}{100\%},
    \label{seq:subopt}\\[2mm]
    & \underline{\smash{\text{UC constraints}}}
    \notag\\
    & \sum_{g \in \mathcal{G}} p^{(i)}_{gt} = L^{(i)}_t
    \qquad  \quad \ \ \forall t \in\mathcal{T},\label{seq:de1}
    \\
    & \bigl(\bf{p}_g^{(i)},\bf{u}_g^{(i)}\bigr) \in \Pi_g^{(i)}
    \qquad  \forall g \in \mathcal{G}, \label{seq:tau_g_polytype1}
    \end{align}
\end{subequations}
where $d(\cdot)$ is a function that measures the tightness of the thresholds. Constraints \eqref{seq:tau_minus_toggle}--\eqref{seq:tau_plus_toggle} enforce the fixing rule: probabilities above $\oli{\tau}_g$ require $u_{gt}^{(i)}=1$, while probabilities below $\uli{\tau}_g$ require $u_{gt}^{(i)}=0$. Constraint \eqref{seq:pos} ensures that the thresholds are valid and ordered within the probability range. The suboptimality constraint \eqref{seq:subopt} ensures that the resulting UC solution does not exceed the user-defined cost tolerance relative to the optimal value $C^{*(i)}$. The remaining constraints (i.e., \eqref{seq:de1}--\eqref{seq:tau_g_polytype1}) correspond to the UC formulation for instance~$i$.

\begin{figure}[t]
    \centering
    \vspace{0mm}
    \includegraphics[width=\linewidth]{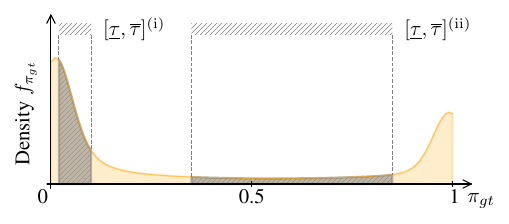}
    \vspace{-8mm}
    \caption{Effect of the threshold interval location and width on variable fixing. The narrower interval $[\uli{\tau}_g,\oli{\tau}_g]^{\rm{(i)}}$ contains more probability mass than the wider interval $[\uli{\tau}_g,\oli{\tau}_g]^{\rm{(ii)}}$, therefore leaving a larger share of variables free.}
    \label{fig:proba_dist}
\end{figure}

An obvious choice for the tightening function $d(\cdot)$ would be to simply minimise the total geometric width of the threshold intervals, i.e., choosing
\begin{equation}\label{eq:width_obj}
    d\left(\uli{\bs{\tau}},\oli{\bs{\tau}}\right)=\sum_{g \in \mathcal{G}} \oli{\tau}_g-\uli{\tau}_g\,.
\end{equation}
However, this approach neglects any information about the distribution of the predicted probabilities $\pi_{gt}$ that can be used to control the fixing behaviour. We illustrate this with Figure \ref{fig:proba_dist}, showing that tighter thresholds do not necessarily lead to a higher proportion of fixed variables. Instead, we may aim to minimise the probability mass of $\pi_{gt}$ that is covered by the interval. This corresponds to minimising:
\begin{equation}\label{eq:density_obj}
    d\left(\uli{\bs{\tau}},\oli{\bs{\tau}}\right)
    =
    \sum_{g\in\mathcal{G}}\sum_{t\in\mathcal{T}}
    \int_{\uli{\tau}_g}^{\oli{\tau}_g}
    f_{\pi_{gt}}(x)\,\mathrm{d}x.
\end{equation}
Here, $f_{\pi_{gt}}$ denotes the probability density of the predicted commitment probability $\pi_{gt}$ for generator $g$ at time $t$. The thresholds remain generator-specific and are shared across all time periods, while the objective measures the expected number of commitment variables left free.
In its current form, \eqref{eq:density_obj} cannot be passed directly to an off-the-shelf MILP solver, since the probability mass generally depends nonlinearly on the threshold decisions through the integration limits. We therefore use a quantile-based linear approximation, detailed in Appendix~\ref{appendix:a_linearization}.

If the thresholds were tuned jointly over all validation instances, the UC decision variables $\mathbf{P}^{(i)}$ and $\mathbf{U}^{(i)}$, together with constraints \eqref{seq:tau_minus_toggle}--\eqref{seq:tau_g_polytype1}, would need to be copied for every $i\in[1:N^{\text{val}}]$. Since $N^{\text{val}}$ should be sufficiently large to ensure that thresholds are robust under a wide range of operating conditions, the resulting optimisation problem may be intractable. We tackle this computational burden by employing Algorithm \ref{alg:tau_tuning_benders}, which iteratively widens the thresholds.

\begin{algorithm}[h!]
\caption{Decomposed $\tau$-tuning}
\label{alg:tau_tuning_benders}
\begin{algorithmic}[1]
\Require Validation instance parameters $\bigl\{\bs{\Theta}^{(n)}\bigl\}_{n\in[1:N^{\rm{val}}]}$, validation probability scores $\bigl\{\bs{\pi}^{(n)}\bigl\}_{n\in[1:N^{\rm{val}}]}$, suboptimality limit $\varepsilon^{\max}_{\%}$, maximum number of optimality cuts $K^{\max}$ per subproblem

\State $i\gets0$, $\uli{\bs{\tau}}^0\gets\bs{0}_{|\mathcal{G}|}$, $\oli{\bs{\tau}}^0\gets\bs{1}_{|\mathcal{G}|}$, $\Phi^0(\uli{\bs{\tau}},\oli{\bs{\tau}})\gets\mathrm{True}$

\While{not finished}
    \State Solve the master problem:
    \begin{subequations}\label{eq:MP}
        \begin{align}
        \min_{\uli{\bs{\tau}},\oli{\bs{\tau}}} \quad & d\left(\uli{\bs{\tau}},\oli{\bs{\tau}}\right)\notag\\
        \text{s.t.} \quad & 0\le\uli{\tau}_g\le\oli{\tau}_g\le1\quad\forall g\in\mathcal{G}\notag\\
        & \Phi^i(\uli{\bs{\tau}},\oli{\bs{\tau}})\notag
        \end{align}
    \end{subequations}
    \hspace{6mm} Record solution: $\bs{\tau}^{i}\gets\left\{\uli{\bs{\tau}}^*,\oli{\bs{\tau}}^*\right\}$
    \For{$n = 1$ to $N^{\text{val}}$}
    
        \State Check whether constraints \eqref{seq:tau_minus_toggle}--\eqref{seq:tau_g_polytype1} are feasible 
        \Statex \hspace{\algorithmicindent}\hspace{\algorithmicindent}with respect to\ $\bs{\Theta}^{(n)}$, $\bs{\pi}^{(n)}$, $\bs{\tau}^{i}$ and $\varepsilon_{\%}^{\max}$
        \If{infeasible}
            
        \State $\uli{\mathcal{F}}\gets(g,t)$-indices where \eqref{seq:tau_minus_toggle} fixes $u_{gt}$ to 0
        \State $\oli{\mathcal{F}}\gets(g,t)$-indices where \eqref{seq:tau_plus_toggle} fixes $u_{gt}$ to 1
        \State $\phi\gets\textsc{AddCuts}\bigl( \bs{\Theta}^{(n)},\bs{\pi}^{(n)}, \uli{\mathcal{F}},\oli{\mathcal{F}}, \varepsilon_{\%}^{\max},K^{\max}\bigr)$
            \State $\Phi^{i+1}\gets\Phi^{i}\wedge\phi$
            \State $i\gets i+1$
            \State Go to step 3.
        \EndIf
    \EndFor
    \State \textbf{break while}
\EndWhile

\State \Return $\bs{\tau}^{i}$
\end{algorithmic}
\end{algorithm}

The algorithm can be interpreted as a logic-based Benders decomposition approach \cite{hookerLogicBasedBendersDecomposition2024}, where the master problem (MP) iteratively receives feasibility cuts from the infeasible mixed-integer subproblems, which correspond to the UC instances in the validation dataset. Algorithm \ref{alg:tau_tuning_benders} starts with an empty MP, allowing for arbitrarily tight thresholds. The computed thresholds are then tested against all $N^{\rm{val}}$ instances. Upon encountering an infeasible instance, the \textsc{AddCuts} subroutine (Algorithm \ref{alg:generate_cuts}) is solved to find up to $K^{\max}$ candidate cuts on the threshold bounds. The generated conditions $\bigl\{\varphi^1,\ldots,\varphi^{K^{\max}}\bigr\}$, that make the instance feasible, are combined by a logical OR. The higher $K^{\max}$, the more combinations are available for the MP to choose from. While this allows the MP to find overall tighter threshold pairs $[\uli{\bs{\tau}},\oli{\bs{\tau}}]$, it also makes the MP harder to solve. These candidate cuts are then added to the MP via the predicator function $\Phi$ which collects the cuts from all the infeasible subproblems found so far. The logical AND/OR conditions contained in $\Phi$ are modelled via auxiliary binary variables that are added to the MP on the fly.

Upon successful termination, the returned thresholds admit a UC solution within the prescribed cost tolerance for every validation instance. Performance on unseen instances is
evaluated in Section~\ref{sec:exp_study_results}. With a finite set of explored alternatives, the resulting thresholds might not be globally optimal.








\begin{algorithm}[t]
\caption{\textsc{AddCuts}}
\label{alg:generate_cuts}
\begin{algorithmic}[1]
\Require Instance parameters $\bs{\Theta}^{(n)}$, probability scores $\bs{\pi}^{(n)}$, index sets $\uli{\mathcal{F}},\oli{\mathcal{F}}$, maximum number of cuts $K^{\max}$

\State $\phi\gets\mathrm{False}$

\For{$k=1$ to $K^{\max}$}
    \State $\varphi^{k}\gets\rm{True}$
    \State Solve the relaxation problem:
    \begin{subequations}\label{eq:relax}
        \begin{align}
        \min_{\bs{\nu}\in\{0,1\}} \quad & \sum_{(g,t)\in\uli{\mathcal{F}}\cup\oli{\mathcal{F}}}\nu_{gt}\notag\\
        \text{s.t.} \quad & u_{gt}\le \nu_{gt}\quad\forall(g,t)\in\uli{\mathcal{F}}\notag\\
        & 1-u_{gt}\le \nu_{gt}\quad\forall(g,t)\in\oli{\mathcal{F}}\notag\\
        & \sum_{(g,t)\in\mathcal{R}^{\ell}}\nu_{gt}\le|\mathcal{R}^{\ell}|-1\quad\forall\ell\in \{1,\ldots,k-1\}\notag\\
        & \eqref{seq:subopt}-\eqref{seq:tau_g_polytype1}\notag
        \end{align}
    \end{subequations}
    \State Record the indices of the current relaxation:
    \Statex\hspace{4mm} $\mathcal{R}^k\gets\left\{(g,t):\nu_{gt}^*=1\right\}$
    \For{$(g,t)\in\mathcal{R}^k$}
        \If{$(g,t)\in\uli{\mathcal{F}}$}
            $\varphi^{k}\gets \varphi^{k}\wedge\bigl(\uli{\tau}_g\le\pi^{(n)}_{gt}\bigr)$
        \ElsIf{$(g,t)\in\oli{\mathcal{F}}$}
            $\varphi^{k}\gets \varphi^{k}\wedge\bigl(\oli{\tau}_g\ge\pi^{(n)}_{gt}\bigr)$
        \EndIf
    \EndFor
    \State $\phi\gets\phi\vee\varphi^k$
\EndFor

\State \Return $\phi$
\end{algorithmic}
\end{algorithm}

\section{Experimental Setup and Results}
\label{sec:exp_study_results}

\begin{figure*}[t]
    \centering
    \vspace{0mm}
    \ifallknn
        \includegraphics[width=\linewidth]{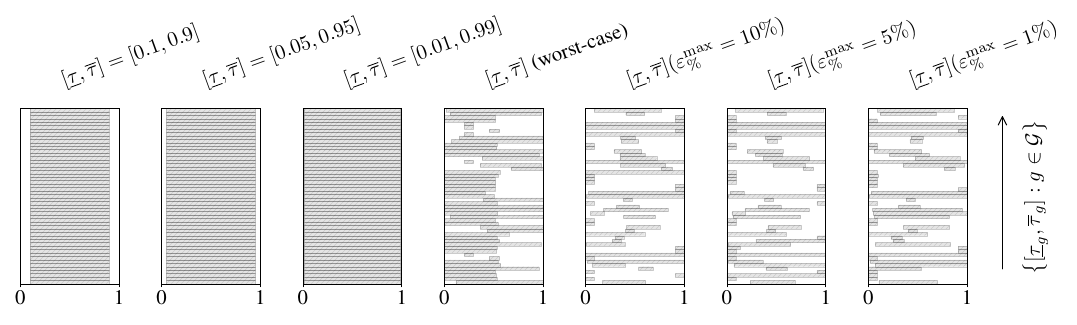}
    \fi
    \caption{Confidence thresholds obtained from hardcoded percentages ($r_{\%}\in\ \{10\%,5\%,1\%\}$), worst-case misprediction, and suboptimality-constrained threshold tuning with $\varepsilon_{\%}^{\max}\in\ \{10\%,5\%,1\%\}$. Each horizontal interval represents one generator, with generators ordered consistently across panels.}
    \label{fig:tau_ranges}
\end{figure*}

This section describes the experimental setup used to evaluate the performance of the various methods introduced in this paper. We first introduce the case study and the dataset. Subsequently, we give an overview of the tested methods, including the ML models used to predict commitment decisions. Finally, we present the results obtained in terms of solution time and solution quality.

\subsection{Case Study}\label{ssec:case_Study}

The proposed framework is evaluated using the dataset from the \textit{AI-ccelerating Unit Commitment} competition \cite{AIcceleratingUnitCommitment}. The test system is based on a simplified representation of the Irish power system under a copper-plate configuration assumption, resulting in a single-bus UC formulation without explicit transmission constraints. The system includes 51 thermal generating units, 15 hydro plants, 13 battery storage units, one pumped-storage hydro unit, and aggregate representations of wind and solar generation. Each UC instance is solved over a 72-hour horizon with hourly resolution. Perfect foresight is assumed for demand and renewable generation availability over the scheduling horizon. Using the default \textit{UnitCommitment.jl} model from \cite{santosxavierUnitCommitmentjlJuliaJuMP2025}, every instance contains 19,296 binary variables, 14,328 continuous variables, and 44,327 constraints. The full dataset contains 2,621 UC instances. 


The proposed approaches are evaluated against a full MILP benchmark solved directly with Gurobi. Gurobipy was used to replicate a Python version of \cite{santosxavierUnitCommitmentjlJuliaJuMP2025}. The full MILP benchmark is solved with a relative MIP gap tolerance of 0.25\%.  For each test instance, the trained ML model predicts commitment decisions and confidence scores. Depending on the thresholding strategy, a subset of commitment variables is fixed and the resulting reduced UC problem is solved. The performance is evaluated in terms of solution time and accuracy. Solution quality is assessed relative to a common lower bound obtained from the full MILP benchmark, as defined in Section~\ref{ssec:performance_evaluation}.
All experiments were performed on a compute node equipped with an AMD EPYC 7H12 64-Core Processor (8 logical CPUs) and 16 GB RAM, using Python 3.11.3 and Gurobi 11.0.0. 


\subsection{Tested Methods}\label{ssec:ml_architecture}

\ifallknn
    Theoretically, the various binary fixing methods introduced in this paper can be adopted using any ML classifier capable of producing probability scores. To allow for a fair comparison between the different methods we focus on a single ML architecture, namely a $k$NN classifier. This choice is motivated by its popularity in existing works using ML for accelerating the solution of the UC (e.g., \cite{dalalUnitCommitmentUsing2018,xavierLearningSolveLargeScale2021,pinedaLearningUnitCommitment2022,gaoOnlineLearningStable2023}). We use the classifier in two distinct ways:
    
    \begin{enumerate}
        \item  We replicate the method proposed by \cite{pinedaLearningUnitCommitment2022}. Unlike a standard $k$NN classifier that assigns labels via a voting rule, their method identifies the $k$ nearest historical instances, iteratively solves the resulting LPs, and selects the lowest-cost, feasible UC schedule. As such, we later refer to this method as ``cost-ranked''.
        \item Although a $k$NN model is typically used to predict discrete class labels, it can also produce probability scores. Here, these scores are computed using inverse-distance weighting.  Let $\mathcal{N}_k^{(i)}$ denote the indices of the $k$ nearest training instances to instance $i$, and let $\rho_{ij}=\lVert\mathbf{z}^{(i)}-\mathbf{z}^{(j)}\rVert_2$ denote their distances in feature space. Assuming no two instances to have identical features, and hence, $\rho_{ij}>0\,,\,\forall j \in \mathcal{N}_k^{(i)}$, the predicted probability score is: 
            \begin{equation}\label{eq:knn_prob}
                \pi_{gt}^{(i)}
                =
                \frac{
                    \sum_{j\in\mathcal{N}_k^{(i)}}
                    u_{gt}^{*(j)}/\rho_{ij}}
                {
                    \sum_{j\in\mathcal{N}_k(i)}
                    1/\rho_{ij}}.
            \end{equation}

        Using these probabilities, the model can thereby be integrated directly into the methods introduced in Section~\ref{sec:proposed_methodology}: 
        \begin{itemize}
            \item A standard binary classifier (Section~\ref{ssec:class_standard}), using a single threshold of $\tau=0.5$.
            \item A confidence-based classifier with constant grey zones (Section~\ref{sssec:const_thresholds}). To test increasingly conservative fixing behaviour, we consider $(\uli{\bs{\tau}},\oli{\bs{\tau}})=(0.1,0.9)$, $(0.05,0.95)$, and $(0.01,0.99)$.
            \item A confidence-based classifier with grey zones informed by worst-case prediction scores (Section~\ref{sssec:worst_case}).
            \item A confidence-based classifier with grey zones obtained from the proposed tuning routine (Section~\ref{sssec:suboptimal}). Different suboptimality factors are tested, namely $\varepsilon_{\%}^{\max}\in(10\%,5\%,1\%)$. For each model we set $K^{\max}=10$.
        \end{itemize}  
    \end{enumerate}
\fi
\ifallknn
     The classifier is trained with $k=50$, while using the Euclidean distance. The feature vector $\bf{z}$ contains raw features, aggregate system demand, renewable generation, and the initial generator conditions, namely the number of hours each unit has been on or off before the start of the scheduling horizon, and engineered features (collated in Table \ref{tab:input_features}, Appendix~\ref{app:features}). We use a 60\%-20\%-20\% split into training, validation and test datasets.
\fi


Using equation \eqref{eq:knn_prob}, we can compute the probabilities for all commitment variables across all instances in the validation dataset. These probabilities can then be employed for the threshold tuning approaches introduced in Sections \ref{sssec:worst_case} and \ref{sssec:suboptimal}. Figure~\ref{fig:tau_ranges} compares the confidence thresholds for the 51 generators in the test system. The predefined approach uses the same thresholds for all generators, whereas the worst-case misprediction and suboptimality-constrained approaches yield generator-specific thresholds. Their effects on variable fixing, solution quality, and runtime are evaluated next.

\subsection{Performance Evaluation}
\label{ssec:performance_evaluation}


Performance is evaluated on previously unseen UC instances using feasibility rate, solution quality, and total runtime, including inference and downstream optimisation. For each instance $i$, let $DB^{(i)}$ denote the best dual bound obtained from the full MILP and $C^{(i)}_{m}$ the feasible objective value returned by method $m$. The relative
optimality gap then is
\begin{equation}
    \text{Optimality Gap}\,({m,i})
    = \frac{C^{(i)}_{m}-DB^{(i)}}{C^{(i)}_{m}}\times100\%.
\end{equation}
The same bound $DB^{(i)}$ is used for all methods, including the full MILP benchmark. An ML-assisted method may achieve a smaller gap if it improves upon the benchmark's incumbent
solution, since the benchmark is solved to a nonzero optimality tolerance. For instance $i$, given the runtime $T^{(i)}_{\rm{MILP}}$ of the full MILP and the runtime $T^{(i)}_{m}$ of the ML-assisted method $m$, the speedup is calculated as
\begin{equation}
    \text{Speedup}\,({m,i}) = \frac{T^{(i)}_{\rm{MILP}}}{T^{(i)}_{m}}.
\end{equation}

\begin{figure*}[t]
    \centering
    \vspace{0mm}
    \ifallknn
        \includegraphics[width=\linewidth]{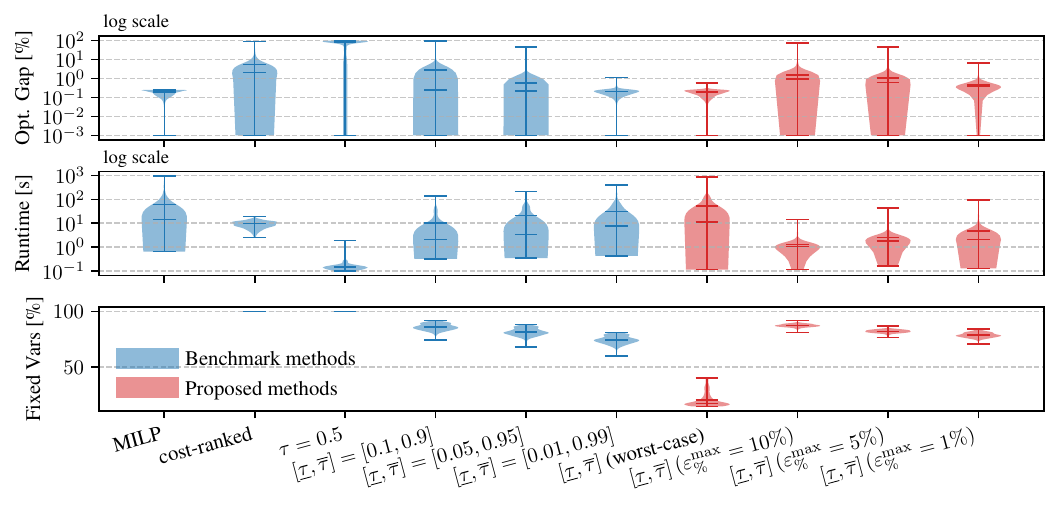}
    \fi
    \vspace{-7mm}
    \caption{Distributions of optimality gap, total runtime, and percentage of fixed binary variables across test instances. Blue denotes benchmark methods and red denotes the proposed methods. For clearer visualisation, the distribution of fixed variables (always zero) for the full MILP was not included.
    }
    \label{fig:method_comparison}
\end{figure*}

The results from the computational study are presented in Figure \ref{fig:method_comparison} and Table \ref{tab:performance_runtime_comparison}. The presented statistics (for the optimality gap, runtime, speedup and fixed variables) are only calculated for the set of feasible instances. Using the computing setup described previously, the full MILP formulation achieved an average solution time of approximately one minute, while the most challenging instance required more than 15 minutes to solve. The following paragraphs briefly discuss the performance of each learning-assisted method.

\begin{table*}[t]
\centering
\caption{Feasibility, optimality, and runtime comparison of different methods on the test dataset}
\label{tab:performance_runtime_comparison}
\begin{tabular*}{\textwidth}{@{\extracolsep{\fill}}lrrrrrrrrr@{}}
\toprule
\multirow{2}{*}{\textbf{Method}} 
& \multirow{2}{*}{\textbf{Feasibility Rate [\%]}} 
& \multicolumn{2}{c}{\textbf{Opt. Gap [\%]}} 
& \multicolumn{2}{c}{\textbf{Runtime [s]}} 
& \multicolumn{2}{c}{\textbf{Speedup [-]}} 
& \multicolumn{2}{c}{\textbf{Fixed Vars [\%]}} \\
& & \multicolumn{1}{c}{mean} & \multicolumn{1}{c}{max} & \multicolumn{1}{c}{mean} & \multicolumn{1}{c}{max} & \multicolumn{1}{c}{mean} & \multicolumn{1}{c}{max} & \multicolumn{1}{c}{mean} & \multicolumn{1}{c}{max} \\
\midrule
\ifallknn
MILP & 100.00 & 0.19 & 0.25 & 60.35 & 947.21 & 1.00 & 1.00 & 0.00 & 0.00 \\
\midrule
cost-ranked & 100.00 & 5.40 & 89.99 & 9.86 & 18.91 & 7.13 & 184.73 & 100.00 & 100.00 \\
\midrule
$\tau=0.5$ & 11.05 & 78.84 & 99.40 & 0.16 & 1.91 & 421.17 & 8175.31 & 100.00 & 100.00 \\
\midrule
$[\underline{\tau}, \overline{\tau}]=[0.1,0.9]$ & 100.00 & 2.73 & 96.04 & 10.16 & 138.56 & 13.88 & 214.68 & 86.10 & 91.94 \\
$[\underline{\tau}, \overline{\tau}]=[0.05,0.95]$ & 100.00 & 0.58 & 46.62 & 20.94 & 215.25 & 6.75 & 91.67 & 81.62 & 88.18 \\
$[\underline{\tau}, \overline{\tau}]=[0.01,0.99]$ & \textbf{100.00} & \textbf{0.21} & 1.13 & 31.25 & 393.99 & \textbf{4.06} & 60.59 & 74.42 & 81.35 \\
\midrule
$[\underline{\tau}, \overline{\tau}]$ (worst-case) & \textbf{99.81} & \textbf{0.20} & 0.59 & 52.90 & 858.13 & \textbf{2.17} & 52.29 & 20.09 & 40.06 \\
\midrule
$[\underline{\tau}, \overline{\tau}]\,(\varepsilon^{\max}_{\%}=10\%)$ & 98.67 & 1.56 & 74.35 & 1.28 & 14.07 & 60.33 & 972.85 & 87.46 & 91.99 \\
$[\underline{\tau}, \overline{\tau}]\,(\varepsilon^{\max}_{\%}=5\%)$ & 98.86 & 1.05 & 45.45 & 2.51 & 43.76 & 34.25 & 1395.66 & 82.25 & 87.12 \\
$[\underline{\tau}, \overline{\tau}]\,(\varepsilon^{\max}_{\%}=1\%)$ & \textbf{99.81} & \textbf{0.48} & 6.63 & 4.70 & 96.28 & \textbf{20.82} & 240.81 & 78.81 & 84.37 \\
\fi
\bottomrule
\end{tabular*}%
\par\vspace{3pt}
\begin{minipage}{\linewidth}
\footnotesize
\textit{Note:} Bold values highlight selected results across the three groups of confidence-based thresholding approaches that achieve the lowest optimality gap within the respective group.
\end{minipage}
\end{table*}

\ifallknn
    \paragraph{Cost-ranked}
    The cost-ranked method maintains feasibility across the entire test set. This behaviour is expected, since it inherently respects individual unit constraints, while selecting from a pool of $k=50$ neighbours provides ample opportunity to identify a system-feasible schedule. Although evaluating 50 LPs in every run increases the total runtime to nearly 19 seconds in the worst case, the average speedup of more than 7\texttimes~still represents a significant acceleration, particularly for challenging instances. The main drawback is its solution quality, with an average optimality gap of more than 5\%. Overall, the results obtained in this study are less favourable than those reported in \cite{pinedaLearningUnitCommitment2022}, where average speedups above 200\texttimes~were achieved while the optimality gap rarely exceeded 0.1\%. We hypothesise that this discrepancy can be attributed to two main factors. First, they consider much larger systems, with up to 6,151 buses, which facilitates higher speedups because of the longer solution times of the full MILP. Second, their UC horizon is limited to 24 hours, compared to the 72-hour horizon considered in this study. Finding a near-optimal schedule may therefore be easier, since prediction errors have less opportunity to propagate and accumulate towards the end of the scheduling horizon.
\fi
 

\paragraph{Standard binary classifier ($\tau=0.5$)}
\ifallknn
    Since this method evaluates only a single complete commitment schedule obtained by hard thresholding at 0.5, it is by far the fastest. However, this speedup comes at the cost of feasibility, as the method produced only 58 feasible solutions across the 525 test instances. Moreover, even when a feasible schedule is found, it is likely to be highly suboptimal.
\fi

\paragraph{Confidence-based fixing (constant thresholds)}
\ifallknn
    Leaving a small number of variables free for the solver to optimise already yields a substantial improvement over hard thresholding at 0.5. Irrespective of $r_{\%}$, these methods consistently yield feasible solutions. The average MIP gap remains below 3\%; yet, in the worst case, it can exceed 90\% and 40\% for the narrower grey zones $[0.1,0.9]$ and $[0.05,0.95]$, respectively. Only the widest grey zone, $[0.01,0.99]$, significantly reduces the worst-case gap. This gain in robustness, however, sacrifices the speedup of some instances, with runtimes exceeding six minutes in the worst case.
\fi

\paragraph{Confidence-based fixing (worst-case misprediction)}
By tuning the thresholds using the probabilities associated with the worst-case mispredictions, this method effectively ``remembers'' specific optimal solutions for which misclassifications would otherwise have occurred.
\ifallknn
    This results in a drastically more conservative fixing behaviour, as illustrated in the lower plot of Figure~\ref{fig:method_comparison}. Although the solution quality remains high across all instances, the average speedup of 2.17 is the lowest among all methods. 
\fi
The results also confirm our suspicion that tighter thresholds may not necessarily fix more variables, motivating the use of a density-based tightening function, as proposed in equation \eqref{eq:density_obj}. While the worst-case method yields tighter thresholds than $[\uli{\tau},\oli{\tau}]=[0.1,0.9]$ (see Figure \ref{fig:tau_ranges}), it fixes significantly fewer variables.

\paragraph{Confidence-based fixing (suboptimality-constrained)}
The suboptimality-constrained variants achieve the most balanced trade-off between feasibility, solution quality, and computational efficiency.
\ifallknn
    Tightening the maximum admissible suboptimality from $\varepsilon^{\max}_{\%}=10\%$ to $1\%$ makes the fixing strategy progressively more conservative, reducing the average proportion of fixed variables from 87.46\% to 78.81\%. This decrease in fixed variables is reflected in an increase of robustness, with $\varepsilon^{\max}_{\%}=1\%$ significantly improving the mean and worst-case optimality gap compared to its less restrictive counterparts. The speedup factors are consistently above $20\times$, outperforming all other confidence-based methods.
\fi
The feasibility rates for the suboptimality-constrained methods are slightly lower, suggesting the existence of a few test instances that would have required a different set of cuts on the thresholds that had not been identified in the validation dataset. After examination, we found that all infeasibilities were due to a violation of the minimum up- and downtime requirements. Augmenting the thresholding procedure with a downstream projection step onto the feasible generator polytope $\Pi_g$ could easily be implemented to prevent this issue.  

Appendix~\ref{app:catboost} evaluates the same fixing strategies using the CatBoost architecture that was used in the competition and examines its performance with respect to speed, solution quality, and feasibility.
\section{Conclusion and Future Work}
\label{sec:conclusions}




\subsection{Concluding Remarks}
This paper revisited confidence-based variable fixing as a means of accelerating the solution of parametric MILP. Using probability scores produced by a ML classifier, we presented three strategies for defining confidence thresholds (constant, worst-case, and suboptimality-constrained) to determine which variables can be fixed with high confidence. For the latter, we proposed a novel optimisation-aware threshold-selection framework that accounts for the impact of fixing errors on the downstream MILP objective through a decomposed threshold-tuning algorithm.
Numerical experiments on a large set of UC instances derived from a realistic power system showed that hard thresholding at 0.5 is unreliable, producing feasible solutions for only a small fraction of the test instances. Constant and worst-case confidence thresholds substantially improve robustness with respect to feasibility and solution quality, but at the expense of achieving lower overall speedups. 
\ifallknn
    The suboptimality-constrained approach yields a favourable trade-off between runtime and solution quality. Depending on the selected tolerance, the method fixed 79--87\% of the binary variables, consistently achieved average speed-ups exceeding 20\texttimes, and maintained mean optimality gaps below 1.6\%, while preserving feasibility for 98.7--99.8\% of the test instances.
\fi
 Overall, the results demonstrate that optimisation-aware confidence calibration provides an effective mechanism for navigating the trade-off between computational speed and solution quality in learning-assisted optimisation.
 
\subsection{Potential Extensions}
While this work focused on finding subsets of binary decisions that can be fixed with high confidence, we envision that the confidence-based thresholding techniques presented in this paper may also be applied to other learning-to-optimise techniques. For instance, warm-starting MILP solvers with variable hints has been employed by numerous works targeting the UC problem~\cite{pourahmadiUnitCommitmentPredictor2025,zaterMultiStageWarmStartDeep2026,omalleyReinforcementLearningMixedinteger2023}. The proposed thresholding could help avoid bad or infeasible warm-starts, which may otherwise have a detrimental effect on runtime rather than improving it. Similarly, ML classifiers for ``constraint-screening'', such as those employed by \cite{pinedaDataDrivenScreeningNetwork2020} for transmission-constrained UC, may benefit from our approach. The classifier could recommend deleting a constraint ($\pi>\oli{\tau}$), retaining it in the model ($\pi\in[\uli{\tau},\oli{\tau}]$), or enforcing it to be binding ($\pi<\uli{\tau}$). Furthermore, our framework could be robustified using predict-and-search methods, such as the approach proposed by \cite{hanGNNGuidedPredictandSearchFramework2023}. Rather than treating the predicted binary decisions as fixed, the search could explore a trust region around the predicted solution by allowing a limited number (e.g., 20) of the ``fixed'' binary variables to change value. Although this introduces an additional hyperparameter, it provides a safeguard against costly misclassifications and can be naturally incorporated into the threshold-tuning procedure described in Section~\ref{sssec:suboptimal}.

Another promising research direction is the further development of the threshold tuning framework itself. In this work, the grey zones $[\uli{\tau},\oli{\tau}]$ are determined during the offline tuning process and subsequently kept fixed. However, a natural extension would be to make the thresholds themselves context-dependent, i.e., $[\uli{\tau}(\bf{z}),\oli{\tau}(\bf{z})]$, allowing them to adapt to the characteristics of each instance. This raises a broader question of whether training the ML classifier $h_{\bs{\theta}}(\bf{z})$ and tuning the thresholds should remain two separate stages of the offline pipeline, or whether they could instead be optimised jointly in an end-to-end learning framework. Beyond adaptive thresholds, another modification of the tuning process could be to explicitly account for uncertainty. Uncertainty sets could be constructed from the historical dataset and incorporated into a robust optimisation framework \cite{guanUncertaintySetsRobust2014}. Rather than optimising the thresholds for the observed training instances alone, one could determine thresholds that satisfy the desired performance criteria for all scenarios within the uncertainty set.

Finally, we emphasise that the proposed framework is not specific to the UC problem but is applicable to a broad class of mixed-integer optimisation problems involving binary decision variables. Promising applications include transmission switching, where binary line status decisions determine the network topology; facility location and network design problems, where binary variables govern infrastructure investments; and scheduling problems, in which binary decisions represent resource assignments or task sequencing.


\appendices
\section{Approximation of Density-Based Objective}
\label{appendix:a_linearization}
We construct a quantile-based linear approximation of the probability-mass objective in \eqref{eq:density_obj}. For this, we observe that the distribution $f_{\pi_{gt}}$ is mainly monotonically decreasing around the lower tail near 0 and monotonically increasing around the upper tail near 1 (cf. Figure \ref{fig:proba_dist}). This observation motivates the following procedure. For each $(g,t)$:
\begin{enumerate}
    \item Collect the set of predicted probabilities $\bigl\{\pi_{gt}^{(i)}\bigr\}_{i\in\mathcal{D}}$, where $\mathcal{D}=\{1,\ldots,N^{\mathrm{val}}\}$ indexes the validation instances.
    \item Divide $\bigl\{\pi_{gt}^{(i)}\bigr\}_{i\in\mathcal{D}}$ into $Q$ quantiles and measure their widths $\bigl\{\Delta_{gt}^1,\ldots,\Delta_{gt}^Q\bigr\}$. We found that $Q=20$ provides a sufficiently accurate approximation for our purpose.
    \item $\forall q \in [1,\ldots,Q]$ create the variables $0\le\Delta\uli{\tau}_{gt}^{q}\le\Delta_{gt}^{q}$ and $0\le\Delta\oli{\tau}_{gt}^{q}\le\Delta_{gt}^{q}$.
    \item Add $\uli{\tau}_{g}\ge\sum_{q=1}^Q\Delta\uli{\tau}_{gt}^q$ and $\oli{\tau}_g\le1-\sum_{q=1}^Q\Delta\oli{\tau}_{gt}^q$ as constraints to the problem.
    \item Find the quantile $q_{0.5}$ above which all probabilities $\pi_{gt}^{(i)}$ are greater than 0.5. We note that this is not the median.
    \item Calculate weights $\uli{w}_{gt}=\{1/\Delta_{gt}^{1},\ldots,1/\Delta_{gt}^{q_{0.5}},0,\ldots,0\}$ and $\oli{w}_{gt}=\{0,\ldots,0,1/\Delta_{gt}^{q_{0.5}+1},\ldots,1/\Delta_{gt}^Q\}$. Let $\uli{w}_{gt}^{q}$ and $\oli{w}_{gt}^{q}$ denote the $q$\textsuperscript{th} entries of these weight vectors. To avoid division by zero and excessively large weights, each reciprocal width $1/\Delta_{gt}^{q}$ is evaluated as $1/\max\{\Delta_{gt}^{q},10^{-6}\}$.
\end{enumerate}
Minimising the probability mass between the thresholds is equivalent to maximising the mass outside them. The resulting linear objective maximises a weighted approximation of this outside-interval mass:

\begin{equation}
    \max\quad
    \sum_{g\in\mathcal{G}}\sum_{t\in\mathcal{T}}\sum_{q=1}^{Q}
    \left(
    \uli{w}_{gt}^{q}\Delta\uli{\tau}_{gt}^{q}
    +
    \oli{w}_{gt}^{q}\Delta\oli{\tau}_{gt}^{q}
    \right).
\end{equation}

The approximation loses its validity in cases where the tails of $f_{\pi_{gt}}$ are not monotonically falling/rising. However, such cases were not encountered in our study.



\section{Input Features}
\label{app:features}

The input vector $\mathbf{z}_t$ is constructed from the raw input data. Let $L_t$ denote the aggregate demand at time $t$, and let $W_t$ and $S_t$ denote the available wind and solar generation, respectively. 
From these, we constructed several new features which are listed in Table \ref{tab:input_features}.

\begin{table}[h!]
\centering
\caption{Input features used by the machine learning models}
\label{tab:input_features}
\renewcommand{\arraystretch}{1.4}
\begin{tabularx}{\columnwidth}{p{0.4\columnwidth} X}
\toprule
\textbf{Feature} & \textbf{Definition} \\
\midrule


RES generation & $RES_t = W_t + S_t$ \\


Net load & $N_t = L_t - RES_t$ \\


Delta for demand, RES generation, net load & $V_t = V_{t}-V_{t-1}, V=\{L,RES,N\}$ \\

3-step moving average of net load &
$\displaystyle \bar{N}^{(3)}_t = \frac{1}{3}\left(N_t+N_{t-1}+N_{t-2}\right)$ \\

3-step rolling maximum of net load &
$\displaystyle N^{(3),\max}_t = \max\{N_t,N_{t-1},N_{t-2}\}$ \\

Mean-normalised net load &
$\displaystyle \frac{N_t}{\bar{N}}, \quad \bar{N}=\frac{1}{|\mathcal{T}|}\sum_{t\in\mathcal{T}}N_t$ \\

Maximum-normalised net load &
$\displaystyle \frac{N_t}{N^{\max}}, \quad N^{\max}=\max_{t\in\mathcal{T}}N_t$ \\

Daily-mean-normalised net load &
$\displaystyle \frac{N_t}{\bar{N}_{d(t)}}, \quad \bar{N}_{d}=\frac{1}{|\mathcal{T}_{d}|}\sum_{t\in\mathcal{T}_{d}}N_t$ \\

Daily-maximum-normalised net load &
$\displaystyle \frac{N_t}{N^{\max}_{d(t)}}, \quad N^{\max}_{d}=\max_{t\in\mathcal{T}_d}N_t$ \\

Time sine harmonics &
$\displaystyle \sin\left(\frac{2\pi t}{24}\right),\quad t\in\mathcal{T}$ \\

Time cosine harmonics &
$\displaystyle \cos\left(\frac{2\pi t}{24}\right), \quad t\in\mathcal{T}$ \\

Initial generator conditions &
$\displaystyle \eta_g^0, \quad g\in\mathcal{G}$ \\
\bottomrule
\end{tabularx}
\end{table}

\section{Alternative Classifier Architectures}
\label{app:catboost}

The original setup used during the competition \cite{AIcceleratingUnitCommitment} was based on an architecture employing CatBoost classifiers~\cite{prokhorenkovaCatBoostUnbiasedBoosting2018} instead of $k$NN models. For completeness, we thus also include the results from using the CatBoost architecture for the same numerical study presented in Section~\ref{sec:exp_study_results}.

\subsection{Model Configuration}
\label{app:hyperparams}

In this case, one CatBoost classifier is trained for each generator. Its predicted probabilities are evaluated across all proposed fixing methods (standard binary, predefined-threshold, worst-case misprediction, and suboptimality-constrained).
The hyperparameters listed in Table~\ref{tab:catboost_params} were obtained via cross-validation and Bayesian optimisation implemented in Optuna~\cite{akibaOptunaNextgenerationHyperparameter07Yue252019}, minimising the average log-loss prediction error compared to the optimal UC schedule of each instance.

\begin{table}[h]
    \centering
    \caption{CatBoost hyperparameters}
    \label{tab:catboost_params}
    \begin{tabularx}{\columnwidth}{p{0.36\columnwidth} X p{0.14\columnwidth}}
    \toprule
    \textbf{Parameter} & \textbf{Description} & \textbf{Value} \\
    \midrule
    \texttt{iterations} & Number of trees & 1475 \\
    \texttt{learning\_rate} & Step size & 0.02935 \\
    \texttt{depth} & Tree depth & 7 \\
    \texttt{l2\_leaf\_reg} & Regularisation & 9.17165 \\
    \texttt{bootstrap\_type} & Weight sampling method & Bernoulli \\
    \texttt{random\_strength} & Split randomness & 1.29014 \\
    \bottomrule
    \end{tabularx}
    \end{table}

\subsection{Results}

Table~\ref{tab:performance_runtime_comparison_catboost} reports the performance of the CatBoost-based fixing methods. The reference statistics of the full MILP are restated for easier comparison. It can be seen that the speed and optimality metrics of the confidence-based methods are improved even further compared to the $k$NN-based models. The feasibility rates are generally lower, since UC schedule predictions are not directly reconstructed from historical solution instances as in the case of $k$NN models.


\begin{table*}[t]
\centering
\caption{Feasibility, optimality, and runtime comparison of different methods on the test dataset}
\label{tab:performance_runtime_comparison_catboost}
\begin{tabular*}{\textwidth}{@{\extracolsep{\fill}}lrrrrrrrrr@{}}
\toprule
\multirow{2}{*}{\textbf{Method}} 
& \multirow{2}{*}{\textbf{Feasibility Rate (\%)}} 
& \multicolumn{2}{c}{\textbf{Opt. Gap (\%)}} 
& \multicolumn{2}{c}{\textbf{Runtime (s)}} 
& \multicolumn{2}{c}{\textbf{Speedup (-)}} 
& \multicolumn{2}{c}{\textbf{Fixed Vars (\%)}} \\
& & \multicolumn{1}{c}{mean} & \multicolumn{1}{c}{max} & \multicolumn{1}{c}{mean} & \multicolumn{1}{c}{max} & \multicolumn{1}{c}{mean} & \multicolumn{1}{c}{max} & \multicolumn{1}{c}{mean} & \multicolumn{1}{c}{max} \\
\midrule
MILP & 100.00 & 0.19 & 0.25 & 60.35 & 947.21 & 1.00 & 1.00 & 0.00 & 0.00 \\
\midrule
$\tau=0.5$ & 1.33 & 30.15 & 80.08 & 0.14 & 0.28 & 425.06 & 6589.06 & 100.00 & 100.00 \\
\midrule
$[\underline{\tau},\overline{\tau}]=[0.1,0.9]$ & 96.00 & 1.50 & 92.70 & 1.08 & 30.94 & 73.04 & 3209.84 & 94.09 & 98.47 \\
$[\underline{\tau},\overline{\tau}]=[0.05,0.95]$ & 99.05 & 0.80 & 90.80 & 2.21 & 33.35 & 35.92 & 563.34 & 91.81 & 96.81 \\
$[\underline{\tau},\overline{\tau}]=[0.01,0.99]$ & \textbf{99.62} & \textbf{0.22} & 3.55 & 10.72 & 161.56 & \textbf{12.54} & 184.60 & 86.49 & 94.09 \\
\midrule
$[\underline{\tau},\overline{\tau}]$ (worst-case) & \textbf{95.62} & \textbf{0.39} & 33.35 & 3.78 & 90.38 & \textbf{36.13} & 2604.34 & 88.01 & 95.45 \\
\midrule
$[\underline{\tau},\overline{\tau}]\,(\varepsilon^{\max}_{\%}=10\%)$ & 94.67 & 0.42 & 5.44 & 1.08 & 19.50 & 82.61 & 3969.15 & 92.75 & 96.08 \\
$[\underline{\tau},\overline{\tau}]\,(\varepsilon^{\max}_{\%}=5\%)$ & 96.00 & 0.41 & 5.09 & 1.16 & 30.28 & 71.68 & 4622.33 & 92.52 & 96.08 \\
$[\underline{\tau},\overline{\tau}]\,(\varepsilon^{\max}_{\%}=1\%)$ & \textbf{98.10} & \textbf{0.33} & 3.74 & 1.50 & 38.64 & \textbf{52.72} & 1559.09 & 84.38 & 87.31 \\
\bottomrule
\end{tabular*}%
\par\vspace{3pt}
\begin{minipage}{\linewidth}
\footnotesize
\textit{Note:} Bold values highlight selected results across the three groups of confidence-based thresholding approaches that achieve the lowest optimality gap within the respective group
\end{minipage}
\end{table*}


\bibliographystyle{IEEEtran}
\bibliography{IEEEabrv,EPRI}






\vfill

\end{document}